\documentclass[aps,prl,twocolumn,superscriptaddress]{revtex4}
\usepackage{graphicx}
\usepackage{dcolumn}
\newcommand{\Dfrac}[2]{\frac{\displaystyle #1}{\displaystyle #2}}
\newcommand{\F}[1]{Figure~\ref{#1}}

\begin{document}

\title{\large Measurement of the $N\to \Delta^+(1232)$ 
              Transition at High Momentum Transfer \\
              by $\pi^0$ Electroproduction }

\newcommand*{\RPI}{Rensselaer Polytechnic Institute, Troy, New York 12180-3590}
\affiliation{\RPI}
\newcommand*{\UCONN}{University of Connecticut, Storrs, Connecticut 06269}
\affiliation{\UCONN}
\newcommand*{\JLAB}{Thomas Jefferson National Accelerator Facility, Newport News, Virginia 23606}
\affiliation{\JLAB}
\newcommand*{\ASU}{Arizona State University, Tempe, Arizona 85287-1504}
\affiliation{\ASU}
\newcommand*{\UCLA}{University of California at Los Angeles, Los Angeles, California  90095-1547}
\affiliation{\UCLA}
\newcommand*{\CSU}{California State University, Dominguez Hills, Carson, CA  90747}
\affiliation{\CSU}
\newcommand*{\CMU}{Carnegie Mellon University, Pittsburgh, Pennsylvania 15213}
\affiliation{\CMU}
\newcommand*{\CUA}{Catholic University of America, Washington, D.C. 20064}
\affiliation{\CUA}
\newcommand*{\SACLAY}{CEA-Saclay, Service de Physique Nucl\'eaire, F91191 Gif-sur-Yvette,Cedex, France}
\affiliation{\SACLAY}
\newcommand*{\CNU}{Christopher Newport University, Newport News, Virginia 23606}
\affiliation{\CNU}
\newcommand*{\ECOSSEE}{Edinburgh University, Edinburgh EH9 3JZ, United Kingdom}
\affiliation{\ECOSSEE}
\newcommand*{\EMMY}{Emmy-Noether Foundation, Germany}
\affiliation{\EMMY}
\newcommand*{\FIU}{Florida International University, Miami, Florida 33199}
\affiliation{\FIU}
\newcommand*{\FSU}{Florida State University, Tallahassee, Florida 32306}
\affiliation{\FSU}
\newcommand*{\GWU}{The George Washington University, Washington, DC 20052}
\affiliation{\GWU}
\newcommand*{\ECOSSEG}{University of Glasgow, Glasgow G12 8QQ, United Kingdom}
\affiliation{\ECOSSEG}
\newcommand*{\ISU}{Idaho State University, Pocatello, Idaho 83209}
\affiliation{\ISU}
\newcommand*{\INFNFR}{INFN, Laboratori Nazionali di Frascati, 00044 Frascati, Italy}
\affiliation{\INFNFR}
\newcommand*{\INFNGE}{INFN, Sezione di Genova, 16146 Genova, Italy}
\affiliation{\INFNGE}
\newcommand*{\ORSAY}{Institut de Physique Nucleaire ORSAY, Orsay, France}
\affiliation{\ORSAY}
\newcommand*{\BONN}{Institute f\"{u}r Strahlen und Kernphysik, Universit\"{a}t Bonn, Germany}
\affiliation{\BONN}
\newcommand*{\ITEP}{Institute of Theoretical and Experimental Physics, Moscow, 117259, Russia}
\affiliation{\ITEP}
\newcommand*{\JMU}{James Madison University, Harrisonburg, Virginia 22807}
\affiliation{\JMU}
\newcommand*{\KYUNGPOOK}{Kyungpook National University, Daegu 702-701, South Korea}
\affiliation{\KYUNGPOOK}
\newcommand*{\MIT}{Massachusetts Institute of Technology, Cambridge, Massachusetts  02139-4307}
\affiliation{\MIT}
\newcommand*{\UMASS}{University of Massachusetts, Amherst, Massachusetts  01003}
\affiliation{\UMASS}
\newcommand*{\MOSCOW}{Moscow State University, General Nuclear Physics Institute, 119899 Moscow, Russia}
\affiliation{\MOSCOW}
\newcommand*{\UNH}{University of New Hampshire, Durham, New Hampshire 03824-3568}
\affiliation{\UNH}
\newcommand*{\NSU}{Norfolk State University, Norfolk, Virginia 23504}
\affiliation{\NSU}
\newcommand*{\OHIOU}{Ohio University, Athens, Ohio  45701}
\affiliation{\OHIOU}
\newcommand*{\ODU}{Old Dominion University, Norfolk, Virginia 23529}
\affiliation{\ODU}
\newcommand*{\PITT}{University of Pittsburgh, Pittsburgh, Pennsylvania 15260}
\affiliation{\PITT}
\newcommand*{\RICE}{Rice University, Houston, Texas 77005-1892}
\affiliation{\RICE}
\newcommand*{\URICH}{University of Richmond, Richmond, Virginia 23173}
\affiliation{\URICH}
\newcommand*{\SCAROLINA}{University of South Carolina, Columbia, South Carolina 29208}
\affiliation{\SCAROLINA}
\newcommand*{\UNIONC}{Union College, Schenectady, NY 12308}
\affiliation{\UNIONC}
\newcommand*{\VT}{Virginia Polytechnic Institute and State University, Blacksburg, Virginia   24061-0435}
\affiliation{\VT}
\newcommand*{\VIRGINIA}{University of Virginia, Charlottesville, Virginia 22901}
\affiliation{\VIRGINIA}
\newcommand*{\WM}{College of William and Mary, Williamsburg, Virginia 23187-8795}
\affiliation{\WM}
\newcommand*{\YEREVAN}{Yerevan Physics Institute, 375036 Yerevan, Armenia}
\affiliation{\YEREVAN}
\newcommand*{\deceased}{Deceased}
\altaffiliation{\deceased}
\newcommand*{\NOWUNH}{University of New Hampshire, Durham, New Hampshire 03824-3568}
\newcommand*{\NOWUMASS}{University of Massachusetts, Amherst, Massachusetts  01003}
\newcommand*{\NOWMIT}{Massachusetts Institute of Technology, Cambridge, Massachusetts  02139-4307}
\newcommand*{\NOWGEISSEN}{Physikalisches Institut der Universitaet Giessen, 35392 Giessen, Germany}
\newcommand*{\NOWNONE}{unknown}
\newcommand*{\NOWGWU}{The George Washington University, Washington, DC 20052}
\newcommand*{\NOWVIRGINIA}{University of Virginia, Charlottesville, Virginia 22901}
\newcommand*{\NOWWM}{College of William and Mary, Williamsburg, Virginia 23187-8795}
\newcommand*{\NOWASU}{Arizona State University, Tempe, Arizona 85287-1504}
\newcommand*{\NOWODU}{Old Dominion University, Norfolk, Virginia 23529}

\author {M.~Ungaro} 
\affiliation{\RPI}
\affiliation{\UCONN}
\affiliation{\JLAB}
\author {P.~Stoler} 
\affiliation{\RPI}
\author {I.~Aznauryan} 
\affiliation{\JLAB}
\affiliation{\YEREVAN}
\author {V.D.~Burkert} 
\affiliation{\JLAB}
\author {K.~Joo} 
\affiliation{\UCONN}
\author {L.C.~Smith} 
\affiliation{\VIRGINIA}

\author {G.~Adams} 
\affiliation{\RPI}
\author {M.~Amarian}
\affiliation{\ODU}
\author {P.~Ambrozewicz} 
\affiliation{\FIU}
\author {M.~Anghinolfi} 
\affiliation{\INFNGE}
% ...

\collaboration{The CLAS Collaboration}
\noaffiliation

\begin{abstract}
 We report a new measurement of the exclusive electroproduction reaction  
 $\gamma^*p~\to~\pi^0 p$ to explore the evolution from soft non-perturbative 
 physics to hard processes via the $Q^2$ dependence of the magnetic 
 ($M_{1+}$), electric ($E_{1+}$) and scalar ($S_{1+}$) multipoles in the  
 $N~\to~\Delta$ transition. 9000 differential cross section 
 data points cover $W$ from threshold to $1.4$~GeV/c$^2$, $4\pi$ 
 center-of-mass solid angle, and $Q^2$ from $3$ to $6$ GeV$^2$/c$^2$, 
 the highest yet achieved. 
 It is found that the magnetic form factor $G^*_M$ decreases 
 with $Q^2$ more steeply than the proton magnetic form factor, the ratio
 $E_{1+}/M_{1+}$ is small and negative, indicating strong 
 helicity non-conservation, and the ratio $S_{1+}/M_{1+}$ is negative, 
 while its magnitude increases with $Q^2$. 
\end{abstract}

\maketitle

The $\Delta(1232)$  resonance is the lowest and most prominent baryon 
excitation, and  the $N~\to~\Delta$ transition has served as a prototype 
for testing theoretical models of baryon structure. For electromagnetic 
excitations in which the $\Delta$ decays into a pion and nucleon, the 
transition amplitudes are expressed in terms of multipoles, which for 
the $N\to\Delta$ transition are the magnetic $M_{1+}$, electric 
$E_{1+}$, and scalar $S_{1+}$~\cite{bib:CGLN}. Alternatively, 
the $N\to\Delta$ transition is expressed in terms of form factors $G^*_M$, 
$G^*_E$ and $G^*_C$~\cite{bib:Jones-Scadron}. 

The $Q^2$ dependence of the electromagnetic multipoles in the $N~\to~\Delta$ 
transition is sensitive to the evolution from soft non-perturbative physics 
to hard processes and perturbative QCD. At low $Q^2$, the small quadrupole 
deformation of the nucleon was long ago understood in the framework of the 
quark model, assuming the reaction is dominated by a single spin-flip of a 
constituent quark in a nearly spherical potential where  $M_{1+}$ is 
dominant~\cite{bib:morpurgo, bib:isgur}. The coupling of the pion cloud to 
the quark core and two body exchange currents may also contribute to the 
small values of the $E_{1+}$ and scalar $S_{1+}$ 
multipoles~\cite{bib:Kamalov-Yang, bib:Buch}. At high $Q^2$, helicity 
conservation in pQCD requires $E_{1+} = M_{1+}$.  

This Letter presents the results of a Jefferson Lab (JLab) experiment  
that extends the measurement of the electromagnetic $N\to\Delta$ transition 
to the highest momentum transfer yet achieved, in order to explore the 
transition region between these low and high $Q^2$ regimes. The unpolarized 
differential cross section for exclusive $\pi^0$ electroproduction has been 
obtained in the hadronic mass $W$ from threshold to $1.4$~GeV/c$^2$, in 
four-momentum transfer $Q^2$ from $3$ to $6$~GeV$^2$/c$^2$, and solid angle 
$4\pi$ in the center of mass.  
The quantities  $G^*_M$, $R_{EM}\equiv Re(E_{1+}/M_{1+})$ and 
$R_{SM}\equiv Re(S_{1+}/M_{1+})$, have been extracted from the 
measured cross sections using a unitary isobar model ~\cite{bib:Inna} that 
takes into account all available data for the $\Delta$ and higher $W$ 
resonances from JLab and other laboratories.

In the one-photon-exchange approximation, the four-fold differential cross 
section of $\pi^0$ electroproduction can be factorized as
$$ 
\Dfrac{d^4\sigma}{dWdQ^{ 2}d\Omega^*_{\pi}} =
 \Gamma_v \Dfrac{d^2\sigma}{d\Omega^*_{\pi}},
$$

where $\Gamma_v$ is the virtual photon flux and  
$d^2\sigma / d\Omega^*_{\pi}$ is the center-of-mass differential 
cross section for $\pi$ production by a virtual photon. 


For the present experiment, an electron beam of energy of $5.75$ GeV was 
incident on a 5.0-cm-long liquid hydrogen target. The CEBAF Large 
Acceptance Spectrometer CLAS~\cite{bib:clas} was used to detect the 
scattered electrons and final state protons. Electrons were selected by 
a hardware trigger formed from the coincidence of signals from a threshold 
gas \v{C}erenkov detector and an electromagnetic calorimeter. Multiwire 
drift chambers were used to reconstruct momenta by measuring particle tracks 
in the CLAS toroidal magnetic field. Plastic scintillators were used to 
record particle time of flight from the interaction point to the 
scintillators.  From their known track length, particle velocities were 
computed and masses calculated using the measured momenta. Software analysis
included geometrical and kinematic cuts to eliminate inefficient areas within 
the spectrometer. Backgrounds coming from $\pi^-/e^-$ contamination were 
suppressed using the energy response in the calorimeter and the signal in the
\v{C}erenkov detector. The  $p\pi^0$ final state was identified using a cut on
the reconstructed missing mass ($M_x^2$) of the detected electron and proton. 
\F{fig:mm_bethe} (left) shows the center-of-mass azimuthal angle of the 
proton $\phi^*_p$ versus $M_x^2$. The most prominent feature is the 
Bethe--Heitler radiative tail (BH) associated with elastic scattering. 
Since the BH events  peak at $M_x^2=0$ and lie primarily in the electron 
scattering plane, they were suppressed by suitable cuts in the 
$M_x^2 $ - $\phi^*_p$  plane. \F{fig:mm_bethe} (right) shows the effects 
of the cuts on the $M_x^2 $ distribution.

\begin{figure}[t]
\includegraphics[width=8.6cm, bb=60 125 500 380]{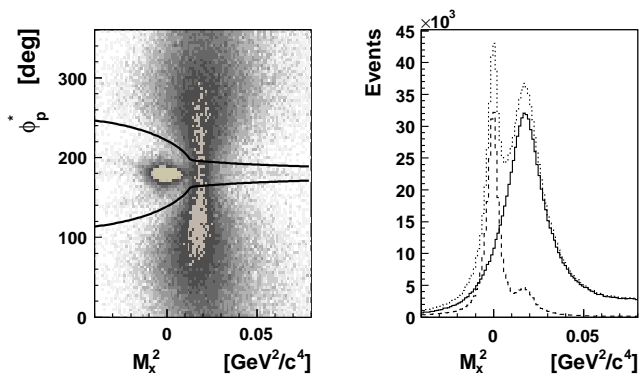}
\caption{The Bethe--Heitler rejection. 
         Left:  $\phi^*_p$ vs $M_x^2$ for $W=1.25$ GeV/c$^2$.  The cuts 
         defined to reject the BH events are shown as solid curves and 
         depend on $W$. Right: the resulting $M_x^2$ distribution in the 
         $W$ region considered. The dotted line shows the $M_x^2$ 
         distribution prior to the cut, the solid line is what remains 
         after the cut, and the dashed line represents the events 
         eliminated by the cut.}
\label{fig:mm_bethe}
\end{figure} 

A Monte Carlo simulation based on GEANT3~\cite{bib:geant} was used to 
determine the acceptance of CLAS and to evaluate the efficiency 
of the BH cuts.  Inelastic radiative losses were corrected 
for using the program EXCLURAD~\cite{bib:radcorr}, which provides a covariant 
treatment of both hard and soft photon radiation in exclusive electroproduction 
and does not rely on the peaking approximation.  

\begin{figure}[b]
\includegraphics[width=8.7cm, bb=70 130 480 590]{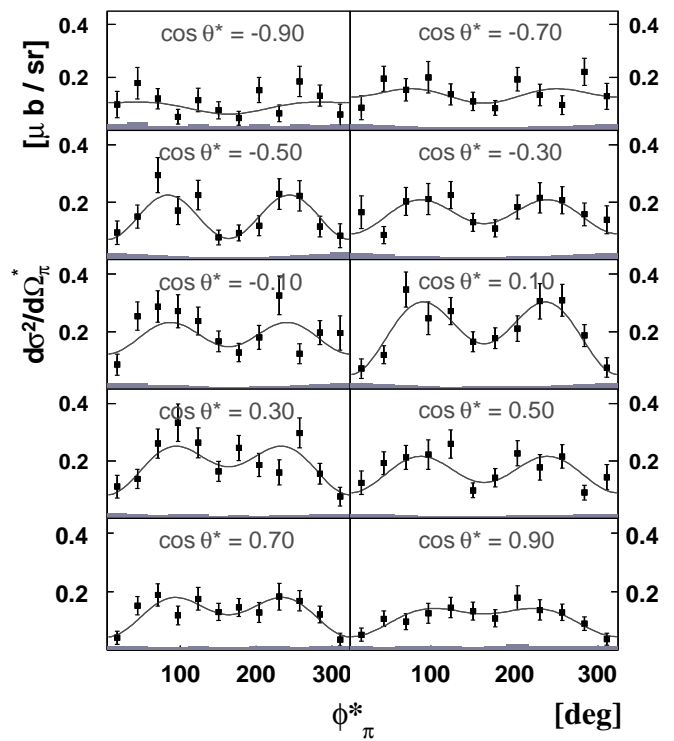}
 \caption{The extracted virtual photon cross section as a function of 
          $\phi^*_{\pi}$  for each $\cos(\theta^*_{\pi})$\ bin in
          the center-of-mass system at $W=1.25$~GeV/c$^2$ and 
          $Q^2=4.2$~GeV$^2$/c$^2$. The error bars are statistical, and 
          the gray band at the bottom of each panel corresponds to the 
          systematic. The solid curves represent the fit using 
          UIM~\cite{bib:Inna}. The fit was carried out utilizing 9000 
          such data points. Each $Q^2$ point was fitted separately. }
\label{fig:sig}
\end{figure}

Differential cross sections were obtained at $9000$ kinematic points, 
binned as follows: $15$ bins in $W$, $5$ bins in $Q^2$, $10$ bins in 
$\cos\theta^*_{\pi}$, and  $12$ bins in $\phi^*_{\pi}$.  Cross 
sections are quoted at the center of each kinematic bin, and a correction 
was calculated to take into account  non-linear dependencies of the cross 
section inside each bin.  Systematic errors were estimated by varying the 
kinematic cuts, such as  $M_x^2$, detector acceptance, particle identification
and vertex reconstruction. Estimated uncertainties in the radiative and bin 
averaging corrections arising from their model dependence are also included. 
Figure \ref{fig:sig} shows an example of the extracted cross sections as a 
function of $\phi^*_{\pi}$ for different $\cos\theta^*_{\pi}$ bins at 
$W= 1.25$~GeV/c$^2$ and $Q^2 = 4.2$~GeV$^2$/c$^2$.

In order to extract the $\Delta$ multipoles $M_{1+}$,  $E_{1+}$
and $S_{1+}$, the truncated multipoles expansion (TME) was commonly 
used at low $Q^2$. In the TME, the structure functions are expanded up 
to $p$- or $d$-waves in Legendre polynomials, whose coefficients are 
related to the multipoles~\cite{bib:raskin}. The magnetic dipole transition 
$|M_{1+}|^2$ is then assumed to dominate the $\pi^0$ production at the 
$\Delta$ pole, and only the terms interfering with $M_{1+}$ are retained.
As the $\Delta$ resonance contribution to the cross section diminishes 
smoothly with increasing  $Q^2$~\cite{bib:Stoler-Phys-Rep}, the TME becomes 
less accurate because $M_{1+}$ dominance is no longer 
assured.  Therefore, models that isolate the $\Delta$ amplitudes from the 
underlying backgrounds must be used.  

The predominantly used approaches have been based on the effective Lagrangian 
expansions, which model the reactions in terms of meson and baryon degrees of 
freedom. MAID~\cite{bib:maid}, which is commonly used to characterize resonance 
amplitudes, is an isobar model approach for photo- and electroproduction data. 
Other elaborations of the effective Lagrangian are the 
Dynamical~\cite{bib:sato-lee}, and DMT~\cite{bib:DMT} models, which couple the 
baryon core and the pion cloud. SAID~\cite{bib:said} is another approach  often 
used to extract amplitudes from global data. 

For the present case, the  unitary isobar model (UIM)~\cite{bib:Inna}, developed 
at JLab, was used. This model incorporates the isobar approach as in 
Ref.~\cite{bib:maid}. The non-resonant background consists of the Born 
term and the {\it t}-channel $\rho$ and $\omega$ contributions. To calculate 
the Born term the latest available measurements of the nucleon and pion form 
factors are used. Underlying tails from resonances such as the $P_{11}(1440)$, 
$D_{15}(1520)$ and $S_{11}(1535)$, which are modeled as Breit--Wigner shapes, 
are also incorporated. The total amplitude is unitarized using the K-matrix 
approach. The dependence of the extracted results on uncertainties in 
non-resonant and higher resonances contributions is included in the systematic 
errors. The results of the fit are given in Table~\ref{tab:result_ratio}.

\begin{figure}[t]
\includegraphics[height=7.9cm, bb=55 140 505 560]{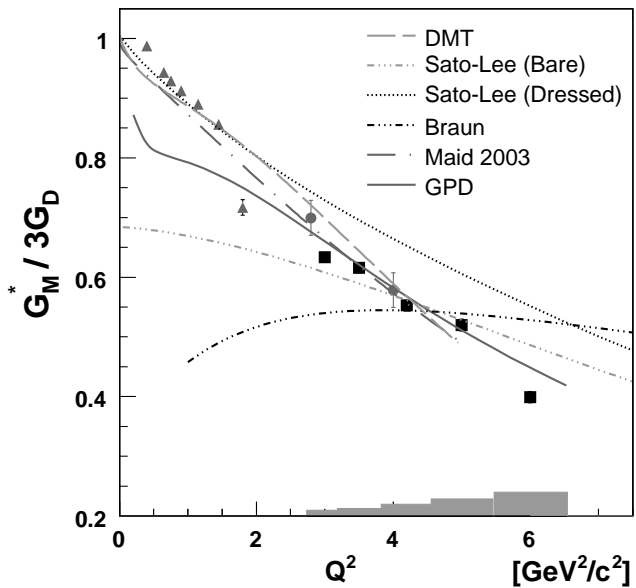}
 \caption{ The form factor $G_M^*/3G_D$. The filled squares are from the 
           current CLAS experiment  utilizing the UIM~\cite{bib:Inna}.  
           The errors shown are statistical, while estimated systematic 
           errors are shown as gray bars at the bottom  of the graph. 
          Also shown are  selected earlier published results. The filled 
          triangles correspond to a recent analysis of previous CLAS 
          data~\cite{bib:colejoo, bib:innaprivate}, and the filled circles are 
          from an earlier JLab Hall C  experiment~\cite{bib:frolov, bib:DMT}. 
          The curves are due to the following calculations. 
          Dashed: dynamical model of Ref~\cite{bib:DMT}. Grey dot-dot-dash: 
          dynamical model of Ref.~\cite{bib:sato-lee} for the ``bare"  
          $\Delta$ (without the pion cloud). Black dotted: full dynamical 
          model of Ref.~\cite{bib:sato-lee}. Black dot-dot-dash:  Light 
          cone sum rule model of Ref.~\cite{bib:Braun-LC-soft}. 
          Dot-dash: MAID-2003~\cite{bib:maid}. Grey solid: GPD model 
          of Ref~\cite{bib:Stoler-Delta}.} 
 \label{fig:GM-JANR}
\end{figure}

Figure \ref{fig:GM-JANR} shows the extracted $G_M^*/3G_D$ as a function 
of $Q^2$ in the Jones--Scadron convention~\cite{bib:Jones-Scadron}. We 
used the 
$M_{1+}~\leftrightarrow~G_M^*$ conversion factor 
$$
 G_M^*={{M_N}\over{ \hbar c k_\gamma}} \sqrt{{{8 p_{\pi}\Gamma_\Delta}
 \over{3\alpha}}\left({1+{{Q^2}\over{(M_\Delta+M_N)^2}}}\right)}
 M_{1+}(M_\Delta),
$$ 
where $k_\gamma$ and $p_{\pi}$ are the center-of-mass momenta of the virtual 
photon and pion respectively, $M_\Delta=1.23$ GeV/c$^2$, the resonance width 
$\Gamma_\Delta$=$120$ MeV, and $G_D=(1+Q^2/0.71)^{-2}$. Also shown are 
selected earlier published results. The most notable feature is that $G_M^*$ 
continues to decrease with $Q^2$ faster than the elastic magnetic form 
factor. This is consistent with Ref.~\cite{bib:goeke-gpd}, which pointed out 
that, through the application of chiral symmetry, $G_M^*$ can be directly 
related to the isovector part of the nucleon elastic form factors. This idea 
was applied in the framework of Generalized Parton Distributions (GPDs) by 
Ref.~\cite{bib:Stoler-Delta}, and later by Ref.~\cite{bib:Guidal-gpd}, to 
suggest that the falloff of $G_M^*$ is related to the falloff of 
$G_E^p$~\cite{bib:Gayou-gep} through their mutual isovector form factor. 

A recent calculation uses the light-cone sum-rules 
(LCSR)~\cite{bib:Braun-LC-soft}. In this approach, the form 
factor is effectively governed  by the overlap of the initial and final QCD 
wave functions. As shown in Fig.~\ref{fig:GM-JANR}, there is modest agreement 
with experiment for $Q^2 $ greater than a few GeV$^2$/c$^2$.

\begin{table}[b]
{\scriptsize
 \begin{center}
  \begin{tabular}{c | c | c | c}        
    \hline\hline
                    &                        &          &           \\
    $Q^2$           & $100 \cdot G_M^*/3G_D$ & $R_{EM}$ &  $R_{SM}$ \\
     GeV$^2$/c$^2$  &                        &  ($\%$)  &   ($\%$)  \\
                    &                        &          &           \\
    \hline\hline
          &                          &                             &                             \\
     3.0  &  $ 63.4\pm 0.2  \pm 0.9$ & $ -1.61  \pm 0.39 \pm 0.22$ &  $-11.5 \pm 0.5  \pm 2.01 $ \\
     3.5  &  $ 61.4\pm 0.4  \pm 1.2$ & $ -1.07  \pm 0.47 \pm 0.10$ &  $-13.0 \pm 0.7  \pm 1.13 $ \\
     4.2  &  $ 55.2\pm 0.5  \pm 1.9$ & $ -3.15  \pm 0.70 \pm 0.20$ &  $-16.4 \pm 1.2  \pm 1.38 $ \\
     5.0  &  $ 52.2\pm 1.0  \pm 2.8$ & $ -3.23  \pm 1.51 \pm 0.33$ &  $-24.8 \pm 2.7  \pm 2.8 $  \\
     6.0  &  $ 39.9\pm 1.5  \pm 4.0$ & $ -3.84  \pm 2.69 \pm 1.40$ &  $-24.8 \pm 5.3  \pm 3.0 $  \\
          &                          &                             &                             \\
    \hline
  \end{tabular}
 \end{center}
 \caption{ Results for $G_M^*/3G_D$, $R_{EM}$ and $R_{SM}$. The first of 
           the quoted errors is statistical, and the second represents our 
           calculation of the systematic uncertainties. The quoted form 
           factor $G_M^*$ is defined according to the Jones--Scadron 
           convention of Ref.~\cite{bib:Jones-Scadron}. 
}
 \label{tab:result_ratio}
}
\end{table}

\begin{figure}[t]
 \includegraphics[height=10.2cm, bb=60 150 510 690]{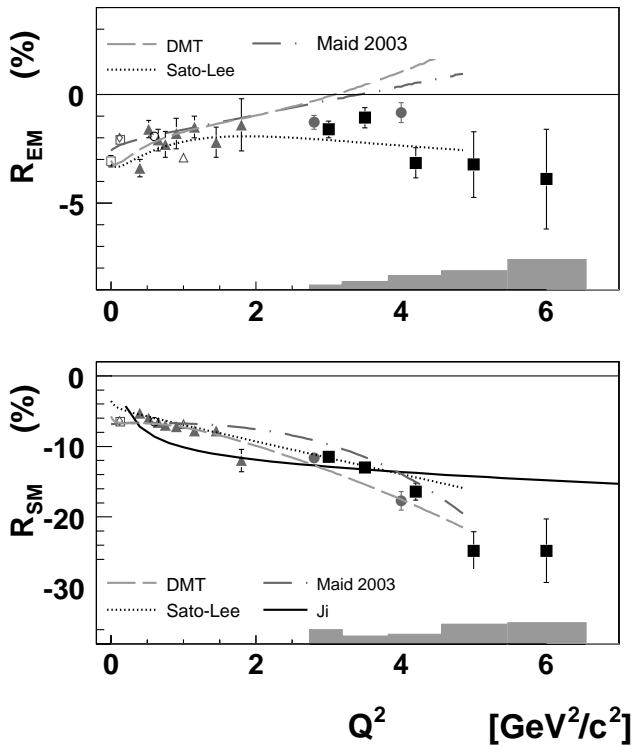}
 \caption{ The ratios $R_{EM}$ (upper panel) and $R_{SM}$ (lower panel).
           The filled red squares are from the UIM fit to the current CLAS 
           experiment. The errors shown are statistical, while estimated 
           systematic errors are shown as gray bars at the bottom of the 
           graph. 
           The filled  triangles at lower $Q^2$ are the previously reported  
           CLAS results~\cite{bib:colejoo}. The filled  circles are from an 
           earlier JLab Hall C  experiment~\cite{bib:frolov, bib:DMT}.  
           The open triangles: JLab Hall-A~\cite{bib:hall-A}. Additional 
           symbols at lower  $Q^2$ are results of measurements  by other 
           laboratories as follows. Open squares: Bates~\cite{bib:bates}. 
           Open trapezoids: Mainz~\cite{bib:mami}. The open circle: 
           Bonn~\cite{bib:elsa}. The curves are as follows. Dotted: full 
           dynamical model of Ref.~\cite{bib:sato-lee}. Dashed: dynamical 
           model of Ref~\cite{bib:DMT}. Dot-dash: MAID-2003~\cite{bib:maid}. 
           Solid: pQCD with orbital angular momentum effects of 
           Ref~\cite{bib:Ji-pqcd}.}
\label{fig:EM-SM-JANR}
\end{figure}

Figure \ref{fig:EM-SM-JANR}  shows the extracted  ratios $R_{EM}$ and 
$R_{SM}$. $R_{EM}$ is small and negative over the entire $Q^2$ range,
indicating strong helicity non-conservation.
$R_{SM}$ is negative and its magnitude increases as a function of $Q^2$.
Our results suggest that the region of $Q^2$ where pQCD processes would be 
expected to be valid is  higher than currently accessible. Adding to the 
controversy, Ref.~\cite{bib:Ji-pqcd} has suggested that pQCD can possibly be 
invoked without strict helicity conservation if orbital angular momentum
flips are included into the perturbative reaction mechanism. The prediction 
for $R_{SM}$ of  Ref.~\cite{bib:Ji-pqcd} is shown in 
Fig.~\ref{fig:EM-SM-JANR} (lower panel).

Progress is being made in describing the $\gamma^* p\rightarrow \Delta$ 
transition at low $Q^2$ using the methods of lattice QCD (LQCD), where 
calculations of the magnetic form factor of this transition are being 
carried out up to $Q^2\sim1.5$ GeV$^2$/c$^2$. The results appear 
encouraging~\cite{bib:LQCD}, however, at the $Q^2$ values of the present 
experiment, the application of LQCD is not yet feasible.  

Included in Figs.  \ref{fig:GM-JANR}  and \ref{fig:EM-SM-JANR} are the 
results of calculations using effective Lagrangian based models whose 
ingredients were tuned to fit earlier data at  lower $Q^2$.  Until a reliable 
treatment in terms of QCD degrees of freedom becomes fully developed, these 
models give unique insights into the baryon structures and their 
manifestations in terms of the traditional hadronic degrees of freedom. 
A review of some of these models and the physical interpretations may be 
found in Ref.~\cite{bib:burk-lee}.

In summary, complete angular distributions for single $\pi^0$ 
electroproduction from protons are reported for a range of $Q^2$ 
from $3$ to $6$  GeV$^2$/c$^2$ and a range of $W$ from $\pi^0$ 
threshold to $1.4$ GeV/c$^2$. The quantities  $G^*_M$, $R_{EM}$, 
and $R_{SM}$ were extracted utilizing the isobar model~\cite{bib:Inna}. 
The results indicate that the form factor $G_M^*$ decreases with $Q^2$ 
faster than the elastic magnetic form factor. $R_{EM}$ is small and 
negative, while $R_{SM}$ remains negative and increases in magnitude. These 
results confirm the absence of pQCD scaling at these kinematics and suggest 
large helicity non-conservation. They provide strong constraints on 
isobar-based effective Lagrangian models, or on approaches employing 
fundamental partonic degrees of freedom such as LQCD, GPDs, LCSR and 
eventually pQCD. However, greater theoretical progress will be 
necessary before good quantitative agreement with the experimental 
high-$Q^2$ data is obtained.

We acknowledge the efforts of the staff of the Accelerator and the Physics 
Divisions at JLab that made this experiment possible. This work was supported 
by the U.S. Department of Energy, the National Science Foundation, the  
Italian Istituto Nazionale di Fisica Nucleare, the French Centre National 
de la Recherche Scientifique and the Commissariat \`{a} l'Energie Atomique, 
and the Korea Science and Engineering Foundation. The Southeastern 
Universities Research Association (SURA) operates the Thomas Jefferson 
National Accelerator Facility for the United States Department of Energy 
under contract DE-AC05-84ER40150.

\end{document}